# EV Charging Station Wholesale Market Participation: A Strategic Bidding and Pricing Approach

Mohammad Mousavi
School of Electrical, Computer and Energy Engineering
Arizona State University
Tempe, AZ, USA
mmousav1@asu.edu

Li "Lisa" Qi
US Research Center ABB Inc
Raleigh, NC, USA
lisa.qi@us.abb.com

Alexander Brissette
US Technology Center ABB E-Mobility
Raleigh, NC, USA
alex.brissette@us.abb.com

Meng Wu
School of Electrical, Computer and Energy Engineering
Arizona State University
Tempe, AZ, USA
mwu@asu.edu

*Abstract*—This paper presents a framework for simultaneous bidding and pricing strategy for wholesale market participation of electric vehicle (EV) charging stations aggregator. The proposed framework incorporates the EV charging stations' technical constraints as well as EV owners' preferences. A bi-level optimization is adopted to model the problem. In the upper level, the total profit of the EV charging station aggregator is maximized. In the lower-level problem, the EV owner's utility function is maximized. The EV owners' preferences are modeled using the quadratic utility function. The bi-level optimization problem which is non-convex and hard to solve is converted to a mixed-integer convex quadratic programming model by writing the optimal conditions of the lower-level problem that is solvable with commercial solvers. The effectiveness of the proposed framework is investigated by implementing simulation results.

## Nomenclature

**Sets and Indices**
$t, T$      Index and set of time slots.
$k, K$     Index and set of charging stations.

**Parameters and Constants**
$\pi_t^e$      Wholesale energy price at time t.
$E_{initial,k}$   Initial charge level of EV of charging station $k$.
$\eta_k^{di}$      Discharging efficiency of EV of charging station $k$.
$\eta_k^{ch}$     Charging efficiency of EV of charging station $k$.
$\underline{E_k}$      Minimum charge level of EV of charging station $k$.
$\overline{E_k}$      Maximum charge level of EV of charging station $k$.
$\overline{DR_k}$    Maximum discharging rate of EV of charging station $k$.
$\overline{CR_k}$    Maximum charging rate of EV of charging station $k$.
$P_k^{Cap}$    Hosting capacity of the distribution network at charging station $k$.

**Variables**
$P_t^{WM}$    Total injected power to wholesale market at time t.
$E_{t,k}$     Charge level of EV of charging station $k$ at time t.
$P_{t,k}^{CS}$     Injected power of charging station $k$ at time t.
$P_{t,k}^{CS,di}$    Discharging power component of the injected power of the charging station $k$ at time t.
$P_{t,k}^{CS,ch}$    Charging power component of the injected power of charging station $k$ at time t.
$b_{t,k}$      Binary variable indicating charging or discharging mode of EV of charging station $k$ at time t.
$\lambda_{t,k}^{EV}$     Energy price bought/sold from/to the EV of charging station $k$ at time t.

## I. INTRODUCTION

The installed capacity of distributed energy resources (DERs) is growing rapidly due to low carbon emissions and increasing electricity demand. Thanks to their low operational costs, DERs can participate in the wholesale energy market and make a profit [1]-[2]. Federal Energy Regulatory Commission issued Order No. 2222, which removes barriers that prevent DERs from wholesale market participation. Based on this order, all independent system operators (ISOs) should revise their tariffs such that DERs with a capacity of greater than 100 kW can participate in the wholesale energy and ancillary service markets [3].

Many works have been focused on the market participation of electric vehicles (EVs) and EV charging stations [4]-[12]. In [4], a dynamic programming approach is used to develop an algorithm for market participation of a load aggregator to manage the charging of the EVs. The load aggregator submits the demand bids to the market. The EVs are modeled by considering a load aggregator which manages the charging of EVs. In [5], a framework is proposed for the participation of the EV aggregators in the wholesale energy and pay-as-bid reserve markets. In [6], a stochastic-based optimal bidding strategy is proposed for wholesale energy and regulation market participation of an EV aggregator. A new battery life cycle modeling is proposed to model the battery characteristics. In [7], an optimization algorithm is proposed for determining optimal energy and reserve bids for EVs aggregated by a market agent. In [8], a bidding strategy is proposed for an aggregator that participates in the wholesale energy and reserve market while compensating the EV owners for degradation. The aggregator is defined as a mediator that trades with the ISO and EV owners. The aggregators maximize their profit for wholesale energy and regulation market participation. In [9], a two-stage stochastic programming approach is proposed for an EV aggregator to participate in the energy and frequency constrained reserve market. Uncertainty of the electricity prices and EV availability is considered. In [10], a coordinated framework for the operation of the EV charging stations and distribution system for providing energy and reserve is proposed based on a cooperative game. In [11], a two-stage stochastic programming model is proposed for market participation of the DERs including the EV charging stations through the distribution system operator which runs the retail electricity market and gathers the offers for wholesale market participation. In [12], the pricing problem of an EV charging station service provider considering uncertainty is proposed. Customer satisfaction and impact on the power grid are modeled. The problem is modeled using multi-objective optimization.

To the best of our knowledge, a *simultaneous* strategic bidding and pricing approach for wholesale market participation of EV charging stations aggregator considering the utility of the EVs owners has not been studied yet. Existing works considered the strategic wholesale bidding and retail pricing problems of EV charging stations *separately* without studying their joint impact on the overall profit of the EV charging stations. Besides, existing works also ignored the interactions between the EV charging station's profit maximization process (achieved by strategically determining the wholesale bids and retail charging prices). In this work, a framework is proposed for simultaneous bidding strategy and pricing problem of the EV charging station aggregator. A bi-level optimization framework is adopted to model the proposed problem. In the upper-level problem, the total profit of the EV charging stations aggregator is maximized. In the lower-level problem, the utility function of the EV owners is maximized. The proposed non-convex and hard-to-solve problem is transformed into the single-level convex and solvable optimization problem by writing the optimality conditions of the lower-level problem. The simulation results are performed to verify the effectiveness of the proposed model.

## II. EV CHARGING STATION MODELING

In this section, the EV charging stations aggregator profit maximization problem is proposed. The EV charging stations aggregator trades with the wholesale market on one side and coordinates the charging of individual EVs on the other side. The EV charging stations aggregator can either buy energy from the EVs and sell it to the wholesale market or buy energy from the wholesale market and sell it to the EVs. The EV charging stations are located in the distribution system which is usually owned by distribution companies. They need to interconnect to the distribution companies to be able to participate in the wholesale market. Some distribution companies provide a guideline for the interconnection to their network and determine a hosting capacity for the location of interconnection for the DERs [13]. Hence, the EV charging stations aggregator does not need to consider distribution network constraints in modeling the problem. The EV charging stations aggregator is going to maximize the total profit considering that the EV owners are going to maximize their utility as well. Hence in the maximization problem of the EV charging stations aggregator, there is a constraint representing the maximization of the utility of the EV owners. Hence, the resulted problem is a bi-level optimization problem. To model the EV owner's utility maximization, we need to know about the utility function of the EV. The general quadratic utility function is as follows:

$$U(w) = aw - bw^2 \qquad (1)$$

where w is the commodity that the investor cares about; $a$ and $b$ are the coefficients of linear and quadratic terms, respectively.

Here, we have considered the quadratic utility function for representing the preference of the EV owners. We need to assume the commodity that EV owners are going to maximize. We have considered two cases: 1) the EV owner cares about the injected power of the EV at each hour ($P_t^{EV}$); 2) the EV owner cares about the final charge level of the EV at the end of the period ($\sum P_t^{EV}$). Let's first start with considering the first case ($w = P_t^{EV}$). The EV charging station profit maximization problem is modeled as follows:

$$\text{Max} \quad Profit = \sum_t [P_t^{WM} * \pi_t^e + \sum_{k \in K} \lambda_{t,k}^{EV} P_{t,k}^{CS}] \qquad (2)$$

s.t.

$$P_t^{WM} = \sum_{k \in K} P_{t,k}^{CS} \quad \forall t \in T \qquad (3)$$

$$E_{1,k} = E_{initial,k} - \frac{1}{\eta_k^{di}} P_{1,k}^{CS,di} + \eta_k^{ch} P_{1,k}^{CS,ch} \quad \forall t \in T, \quad (4)$$

$$\forall k \in K$$

$$E_{t,k} = E_{t-1,k} - \frac{1}{\eta_k^{di}} P_{t,k}^{CS,di} + \eta_k^{ch} P_{t,k}^{CS,ch} \qquad (5)$$

$$\forall t \in T, \forall k \in K$$

$$P_{t,k}^{CS} = P_{t,k}^{CS,di} - P_{t,k}^{CS,ch} \quad t \in T, \forall k \in K \qquad (6)$$

$$\underline{E}_k \leq E_{t,k} \leq \overline{E}_k \quad \forall t \in T, \forall k \in K \qquad (7)$$

$$0 \leq P_{t,k}^{CS,di} \leq b_{t,k} \overline{DR}_k \quad \forall t \in T, \forall k \in K \qquad (8)$$

$$0 \leq P_{t,k}^{CS,ch} \leq (1 - b_{t,k}) \overline{CR}_k \quad \forall t \in T, \forall k \in K \qquad (9)$$

$$0.5 \overline{E}_k \leq E_{initial,k} + \sum_{t \in T} [P_{t,k}^{CS,ch} \eta^{ch} - P_{t,k}^{CS,di}/ \qquad (10)$$

$$\eta^{di}] \leq \overline{E}_k \quad \forall t \in T, \forall k \in K$$

$$P_t \leq P_k^{Cap} \quad \forall t \in T, \forall k \in K \qquad (11)$$

$$\text{Max} U_k(P_{t,k}^{CS}) - \lambda_{t,k}^{EV} P_{t,k}^{CS} = a \sum_{t \in T} P_{t,k}^{CS} - b(P_{t,k}^{CS})^2 - \qquad (12)$$

$$\lambda_{t,k}^{EV} P_{t,k}^{CS}, \forall k \in K$$

The objective function (2) maximizes the total profit of the EV charging stations aggregator which is composed of two terms. The first term is the profit from trading with the wholesale market and the second term is the profit from trading with the EVs. Note that these terms can be negative which means that the EV charging stations aggregator needs to pay. Equation (3) aggregates the power from all EV charging stations. Equations (4)-(5) define the charge level of the EV based on charging and discharging power. Equation (6) defines the injected power based on charging and discharging power. Constraint (7) limits the charge level of the EV based on its minimum and maximum value. Constraints (8) and (9) limit the charge/discharge power of the EV charging station based on the charging/discharging rates. Constraint (10) ensures that EVs are charged more than 50% at the end of the period. Constraint (11) limits the injected/consumed power of the EV charging station to the network with respect to the hosting capacity. Constraint (12) maximizes the utility function of EVs.

Note that there binary variables in Equations (8) and (9) can be eliminated once efficiany ($\eta_k^{ch}, \eta_k^{di}$) is not equal to 1 since the charging and discharging states do not happen at the same time as there is a loss for charging and discharging states [14].

The proposed optimization problem is a bi-level optimization problem. There are two sources of nonlinearities in the above formulation: 1) the term $\sum_{k \in} \pi_{t,k}^{EV} P_{t,k}^{CS}$ in the objective function; 2) Constraint (13).

The lower-level problem is an unconstrained optimization problem. For simplicity let's drop $k$. By writing the partial derivative of the function with respect to $P_t^{CS}$:

$$\frac{\partial [U(P_t^{CS}) - \lambda_t^{EV} P_t^{CS}]}{\partial P_t^{CS}} = a - 2bP_t^{CS} - \lambda_t^{EV} = 0 \quad (13)$$

$$\lambda_t^{EV} = a - 2bP_t^{CS}$$

The new objective function is:

$$\begin{aligned} Profit &= \sum_{t \in T}[P_t^{WM} * \pi_t^e + (a - 2bP_t^{CS})P_t^{CS}] \\ &= \sum_{t \in T}[P_t^{WM} * \pi_t^e + aP_t^{CS} \\ &\quad - 2bP_t^{CS^2}] \end{aligned} \quad (14)$$

s.t

(3)-(11) \hfill (15)

Hence, the bi-level non-convex optimization problem (2)-(12) is converted to an equivalent single-level convex quadratic optimization problem which is given in (14)-(15).

Now, let's consider case 2 ($w = \sum P_t^{CS}$) and derive the formulation. Again, let's drop $k$ for simplicity. The lower-level problem is as follows:

$$\begin{aligned} U\left(\sum_{t \in T} P_t^{CS}\right) - \lambda_t^{EV} P_t^{CS} \\ = a \sum_{t \in T} P_t^{CS} - b(\sum_{j \in T} P_j^{CS})^2 \\ - \lambda_t^{EV} P_t^{CS} \end{aligned} \quad (16)$$

$$\begin{aligned} \frac{\partial [U(\sum_{t \in T} P_t^{CS}) - \lambda_t^{EV} P_t^{CS}]}{\partial P_t^{CS}} \\ = a - 2bP_t^{CS} - 2b \sum_{t \in T, j \neq t} P_t^{CS} \\ - \lambda_t^{EV} = 0 \end{aligned} \quad (17)$$

$$a - 2bP_t^{CS} - 2b \sum_{t \in T, j \neq t} P_t^{CS} = \lambda_t^{EV} \quad (18)$$

$$\sum_{t \in T} \lambda_t^{EV} P_t^{CS} = a \sum_{t \in T} P_t^{CS} - 2b\left(\sum_{j \in T} P_j^{CS}\right)^2 \quad (19)$$

Then, the objective function of the EV charging stations aggregator can be calculated as follows:

$$Profit = \sum_{t \in T}[P_t^{WM} * \pi_t^e] + a \sum_{t \in T} P_t^{EV} \quad (20)$$
$$- 2b\left(\sum_{j \in T} P_j^{EV}\right)^2$$

s.t.

(3)-(11) \hfill (21)

The only term that is nonlinear is the third term. If we prove that this term is convex, the total objective function will be convex.

$$\left(\sum_{j \in T} P_j^{EV}\right)^2 = [P_1 \quad P_2 \quad \dots \quad P_t] \begin{bmatrix} 1 & \cdots & 1 \\ \vdots & \ddots & \vdots \\ 1 & \cdots & 1 \end{bmatrix} \begin{bmatrix} P_1 \\ P_2 \\ \vdots \\ P_t \end{bmatrix} \quad (22)$$

In order to prove that this term is convex, we need to calculate the eigenvalues of the coefficient matrix A:

$$A = \begin{bmatrix} 1 & \cdots & 1 \\ \vdots & \ddots & \vdots \\ 1 & \cdots & 1 \end{bmatrix} \quad (23)$$

$$|A - \lambda I| = 0$$

$$\lambda = t, 0,0 \dots, 0$$

The Matrix A is positive semidefinite. Hence the quadratic function is convex.

## III. SIMULATION RESULTS

In this section, simulation results are implemented on a test case including an EV charging station and an EV. Future work will be carried out on multiple charging stations and EVs. In order to implement simulation results, the following parameters are assumed: the maximum capacity of the EV is considered to be 600 kW. The initial charge level of the EV is assumed to be 30% of the capacity. The charging/discharging rates are assumed to be the same as the capacity, which means that the

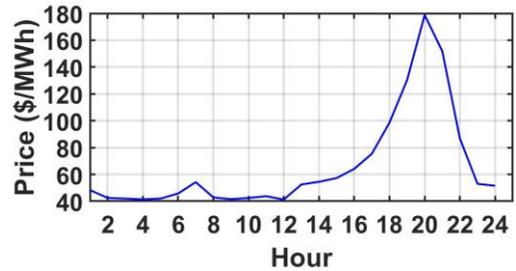

EV can be fully charged or discharged at each hour. The minimum capacity of the EV is assumed to be 20% of the EV's capacity. The coefficient of the linear and quadratic terms of the utility function are considered to be 320 $/MWh and 320 $/MW²h, respectively. The wholesale market energy price is shown in Fig. 1. The utility function of the EV is shown in Fig. 2(a).

Fig. 1. Wholesale market energy price.

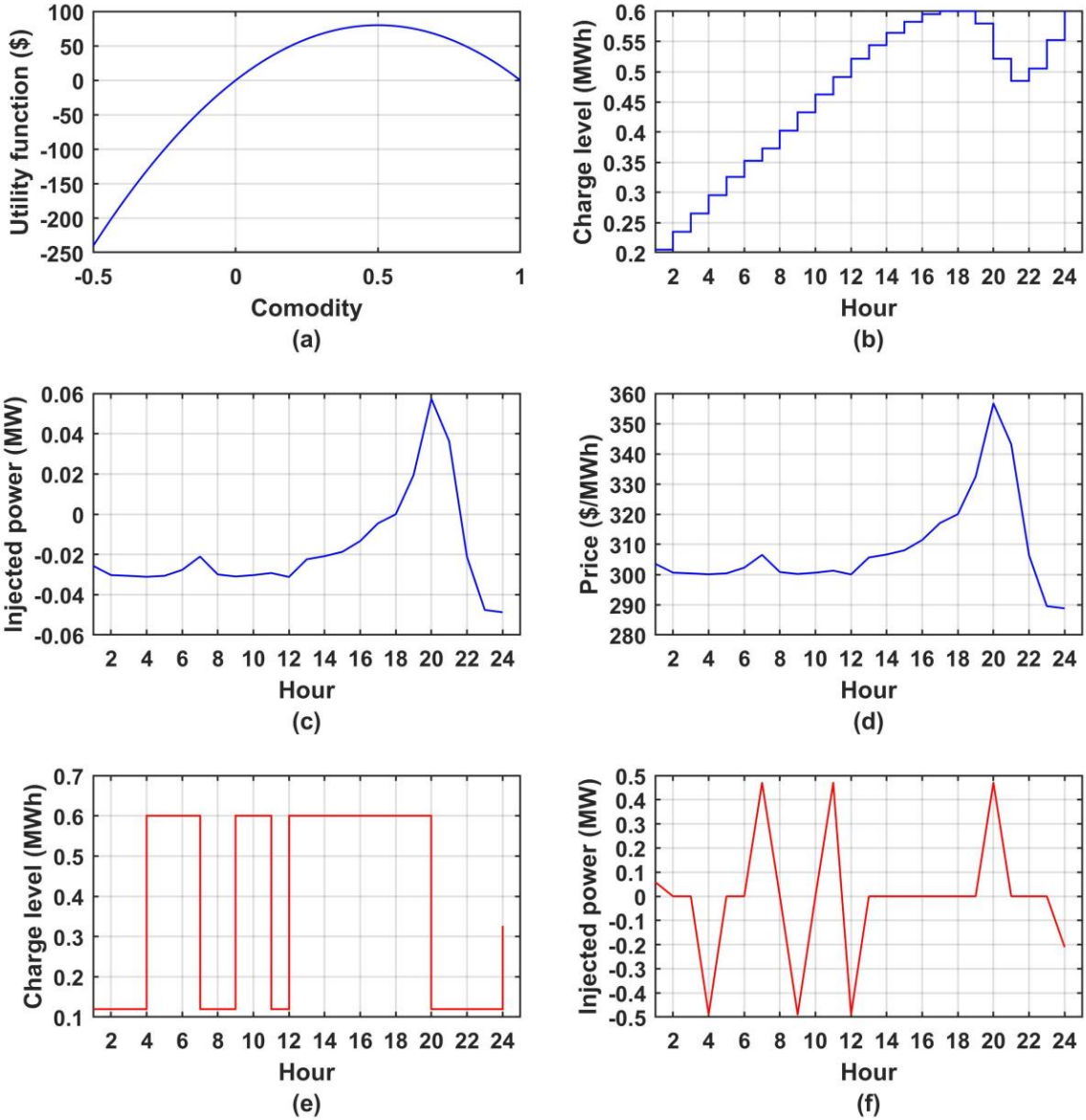

Fig. 2. (a) Utility function of the EV. (b) Charge level of the EV considering the first utility function. (c) Injected power of EV charging station to the network considering the first utility function. (d) Price of trading with EV. (e) Charge level of the considering the second utility function. (f) Injected power of the EV charging station to the network considering second utility function.

*A. First Utility Function*

The charge level of the EV and injected power of the EV charging station are shown in Fig. 2(b) and Fig. 2(c), respectively. It is clear that the injected power of the EV charging station is determined in a way that maximizes the utility function of the EV as well. In the EV objective function, there are two terms. The first term is the profit from trading with the wholesale market and the second term is the profit from trading with EV. At hours 1-18 and 22-24, the wholesale market price is low. Hence, the EV charging station buys from the wholesale market at these hours and charges the EV considering the amount that also maximizes the EV's objective function. However, at hours 19, 20, and 21, the wholesale market price is high and the EV charging station buys energy from the EV and sells it to the wholesale market.

The price of trading with EV at each hour is shown in Fig. 2(d). As expected, it has a shape similar to the wholesale market price.

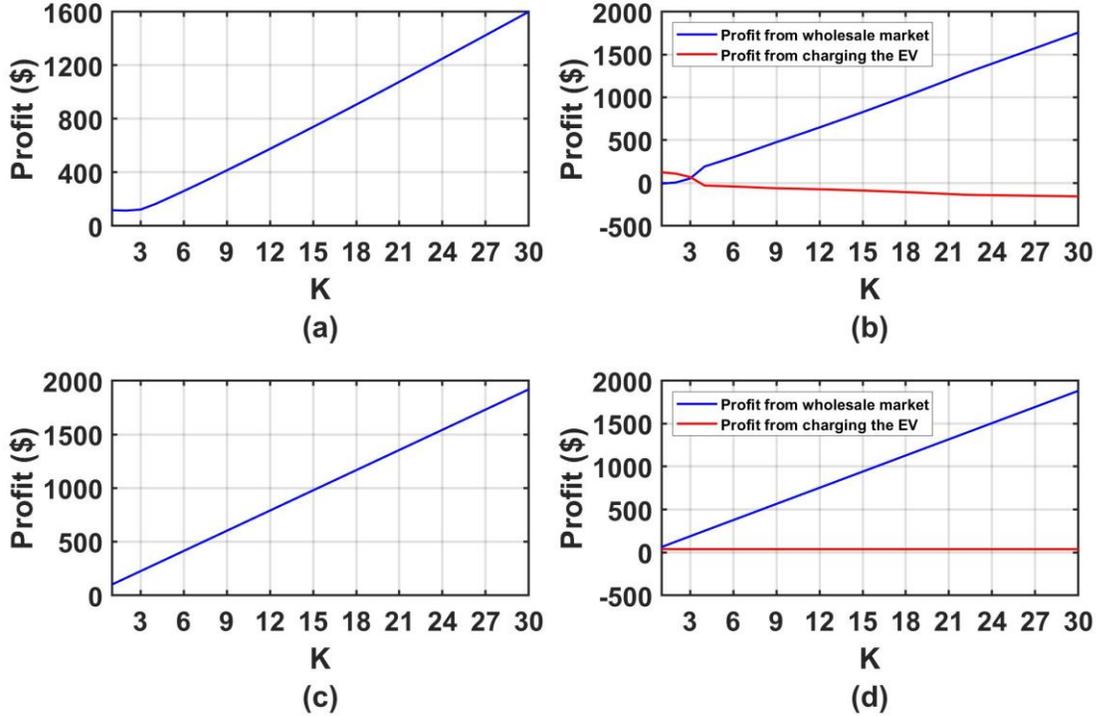

Figure 3. (a) Change in profit with respect to change in wholesale market price using first utility function. (b) Change in profit from wholesale market and charging the EV with respect to change in wholesale market price using first utility function. (c) Change in profit with respect to changes in wholesale market price using the second utility function. (d) Change in profit from the wholesale market and charging the EV with respect to change in wholesale market price using the second utility function.

### B. Second Utility Function

As mentioned, in this scenario, we are considering that the EV owner cares about the final charge level of the EV. The charge level and injected power of the EV charging station to the network in this scenario are shown in Fig. 2(e) and Fig. 2(f), respectively. By looking at the wholesale energy price, one can see that the EV charging station buys energy at hour 4 when the wholesale market price is low and charges the EV and sell it at hour 7 when the wholesale market price is higher than at hour 4. The same thing happens at hours 9 and 11. At hour 12 the EV charging station buys energy from the wholesale market and charges the EV again and saves it until the peak hour which is hour 20 and sells it to the wholesale market. The EV charging station does not sell it at hours between 12 and 20 since it needs to buy it again and the price during these hours is high. The EV charging station buys energy at hour 24, when the price is low again, to charge the EV. The price is 110.51 $/MWh. We expect this flat price since the formulation of the price is not dependent on the $t$.

### C. Sensitivity Analysis

In this section, in order to investigate the effect of the wholesale market prices assumed in the simulation result section, sensitivity analysis is performed.

To perform sensitivity analysis on wholesale market price, all the hourly prices are multiplied by factor $K$ ($\pi_t^{e,new} = K * \pi_t^e$), which changes from 1 to 30, and the problem is solved in each scenario, and profit is calculated.

The effect of changing the wholesale market price on the total profit of the EV using the first utility function is shown in Fig. 3(a). The total profit is approximately constant until K increases to 3. After that, the total profit increases as the wholesale market price increases. In order to investigate the effect of the wholesale market price precisely, the total profit is decomposed to profit from wholesale market participation and the profit from charging the EV. The result is shown in Fig. 3(b). The wholesale market profit increases until K reaches 5 through a curve. The profit from charging the EV decreases as the wholesale market price increases and has a curve shape until K reaches 5. After that, the profit from wholesale market participation increases, and the profit from charging the EV decrease as the wholesale market price increases. However, the effect of profit from the wholesale market exceeds the decreases in profit from charging the EV. Hence, the total profit increase is shown in Fig. 3(a).

The change in total profit with respect to change in the wholesale market price for using the second utility function is shown in Fig. 3(c). The separated profit, including the profit from wholesale market participation and charging the EV, is shown in Fig. 3(d). The change in the total profit and separated profits is linear since the EV charging station only cares about the final charge level of the EV in this case and we have a flat price for EV in this case as discussed earlier.

## IV. Conclusion

In this paper, a simultaneous bidding and pricing strategy approach is proposed for wholesale market participation of EV charging stations aggregators considering the utility function of the EV owners. The EV charging stations aggregator trades with the wholesale market as well as EV owners. The aggregator needs to maximize not only its total profit but also the EV owners' utility function to attract them. The bi-level problem formulation presented in this paper can satisfy the aforementioned needs. In the upper level, the total profit of the EV charging stations aggregator is maximized. In the lower-level problem, the utility function of the EVs is maximized. The bi-level problem which is nonconvex and hard to solve is converted to a mixed-integer convex quadratic programming model by writing the optimality conditions of the lower-level problem which is solvable by commercial solvers. Case studies show that the prices are determined in a way that maximizes the total profit of the EV charging station considering the maximization of the utility of the EV owners.

## References

[1] M. Mousavi and M. Wu, "A DSO Framework for Market Participation of DER Aggregators in Unbalanced Distribution Networks," *IEEE Trans. Power Syst.*, pp. 1–1, 2021, doi: 10.1109/TPWRS.2021.3117571.

[2] M. Mousavi and M. Wu, "A DSO Framework for Comprehensive Market Participation of DER Aggregators," in *2020 IEEE Power Energy Society General Meeting (PESGM)*, Aug. 2020, pp. 1–5. doi: 10.1109/PESGM41954.2020.9281462.

[3] "FERC Opens Wholesale Markets to Distributed Resources: Landmark Action Breaks Down Barriers to Emerging Technologies, Boosts Competition | Federal Energy Regulatory Commission." https://www.ferc.gov/news-events/news/ferc-opens-wholesale-markets-distributed-resources-landmark-action-breaks-down (accessed Nov. 07, 2021).

[4] J. M. Foster and M. C. Caramanis, "Optimal Power Market Participation of Plug-In Electric Vehicles Pooled by Distribution Feeder," *IEEE Trans. Power Syst.*, vol. 28, no. 3, pp. 2065–2076, Aug. 2013, doi: 10.1109/TPWRS.2012.2232682.

[5] C. Goebel and H.-A. Jacobsen, "Aggregator-Controlled EV Charging in Pay-as-Bid Reserve Markets With Strict Delivery Constraints," *IEEE Trans. Power Syst.*, vol. 31, no. 6, pp. 4447–4461, Nov. 2016, doi: 10.1109/TPWRS.2016.2518648.

[6] S. I. Vagropoulos and A. G. Bakirtzis, "Optimal Bidding Strategy for Electric Vehicle Aggregators in Electricity Markets," *IEEE Trans. Power Syst.*, vol. 28, no. 4, pp. 4031–4041, Nov. 2013, doi: 10.1109/TPWRS.2013.2274673.

[7] R. J. Bessa and M. A. Matos, "Optimization Models for EV Aggregator Participation in a Manual Reserve Market," *IEEE Trans. Power Syst.*, vol. 28, no. 3, pp. 3085–3095, Aug. 2013, doi: 10.1109/TPWRS.2012.2233222.

[8] M. R. Sarker, Y. Dvorkin, and M. A. Ortega-Vazquez, "Optimal Participation of an Electric Vehicle Aggregator in Day-Ahead Energy and Reserve Markets," *IEEE Trans. Power Syst.*, vol. 31, no. 5, pp. 3506–3515, Sep. 2016, doi: 10.1109/TPWRS.2015.2496551.

[9] L. Herre, J. Dalton, and L. Söder, "Optimal Day-Ahead Energy and Reserve Bidding Strategy of a Risk-Averse Electric Vehicle Aggregator in the Nordic Market," in *2019 IEEE Milan PowerTech*, Jun. 2019, pp. 1–6. doi: 10.1109/PTC.2019.8810937.

[10] Y. Li, Z. Ni, T. Zhao, Y. Liu, L. Wu, and Y. Zhao, "Coordinated Operation Between Electric Vehicle Charging Stations and Distribution Power Network Considering Shared Energy and Reserve," in *2021 IEEE/IAS 57th Industrial and Commercial Power Systems Technical Conference (I CPS)*, Apr. 2021, pp. 1–10. doi: 10.1109/ICPS51807.2021.9416626.

[11] "A Two-stage Stochastic Programming DSO Framework for Comprehensive Market Participation of DER Aggregators under Uncertainty." https://ieeexplore.ieee.org/abstract/document/9449806/ (accessed Nov. 08, 2021).

[12] C. Luo, Y.-F. Huang, and V. Gupta, "Stochastic Dynamic Pricing for EV Charging Stations With Renewable Integration and Energy Storage," *IEEE Trans. Smart Grid*, vol. 9, no. 2, pp. 1494–1505, Mar. 2018, doi: 10.1109/TSG.2017.2696493.

[13] ComEd, "DER Interconnection Guidelines for Customers Interconnection for Parallel Generation," Jul. 2018. https://www.comed.com/SiteCollectionDocuments/MyAccount/MyService/DER_Interconnection_Guidelines_for_Customers.pdf

[14] L. Bai, J. Wang, C. Wang, C. Chen, and F. Li, "Distribution Locational Marginal Pricing (DLMP) for Congestion Management and Voltage Support," *IEEE Trans. Power Syst.*, vol. 33, no. 4, pp. 4061–4073, Jul. 2018, doi: 10.1109/TPWRS.2017.2767632.